\documentclass[aps,twocolumn,nofootinbib]{revtex4-1}
\usepackage{amssymb}
\usepackage{amsmath}
\usepackage{graphicx}
\usepackage[sort&compress]{natbib}
\bibpunct[, ]{}{}{,}{s}{,}{,}

\usepackage[version=3]{mhchem} 
\begin{document}

\title{Degradation versus self-assembly of block copolymer micelles}

\author{Alexander Muratov, Vladimir A. Baulin$^{\dag}$}
\affiliation{Departament d'Enginyeria Quimica, Universitat Rovira
i Virgili, Av. dels Paisos Catalans 26, 43007 Tarragona, Spain and
$^\dag $ICREA, Passeig Lluis Companys 23, 08010 Barcelona, Spain}

\email{vladimir.baulin@urv.cat}

\begin{abstract}
The stability of micelles self-assembled from block copolymers can be
altered by the degradation of the blocks. Slow degradation shifts the
equilibrium size distribution of block copolymer micelles and change their
properties. Quasi-equilibrium scaling theory shows that the degradation of
hydrophobic blocks in the core of micelles destabilize the micelles reducing
their size, while the degradation of hydrophilic blocks forming coronas of
micelles favors larger micelles and may, at certain conditions, induce the
formation of micelles from individual chains.
\end{abstract}

\maketitle
Journal link:

\url{http://pubs.acs.org/doi/abs/10.1021/la204625p}

\section{Introduction}

Physicochemical properties of block copolymers often determine their
function in many useful applications ranging from biotechnology and drug
delivery to painting and oil extraction\cite{Hamley,Reiss}. The possibility
to control essential properties of block copolymers capable to adapt the
behavior to changes in the environment is thus a challenging task. One of
the main properties of amphiphilic diblock copolymers in solution is their
ability to self-assemble in micelles\cite{Alexandridis} composed of a hydrophobic core
surrounded by a hydrophilic corona\cite{Adams}. Such compact finite size
aggregates can encapsulate hydrophobic agents in their cores\cite{Torchilin1}%
. In particular, this loading capacity of block copolymers can be used for
selective transport of hydrophobic nanoparticles and lipophilic active
molecules to specific targets and through the cell membrane\cite{Savic}.

However, the use of block copolymer micelles for targeted drug delivery
implies also the necessity to control the release of the transported
particles from the cores of micelles, for example by external stimuli. In
turn, the release process is closely related to thermodynamic stability of
micelles. Micelles assembled from block copolymers can be relatively stable
\cite{Husseini}. This hinders the release of active components\cite%
{Torchilin} and thus limits their use for biomedical applications.

Degradable polymers\cite{Scott} provide for additional degree of freedom
allowing to control the longevity and stability of block copolymer micelles.
Degradation of the polymer backbone may change significantly the
thermodynamics of block copolymer self-assembly and thus stability of the
micelles. Tuning the rate of degradation would allow for modulation of the
thermodynamic stability of micelles in large extent. In addition, tumor
tissues have tendency to selectively accumulate polymers. This effect is
known as the enhanced permeation and retention (EPR)\cite{Maeda} and is
attributed to larger size of the pores in blood vessels of tumor tissues.
Polymers can interact with the cell membranes and reticuloendothelial system
(RES)\cite{Stayton}, what can potentially increase their cytotoxicity\cite%
{Slater}. From this perspective, the degradation of polymers up to
metabolites\cite{Bastioli} may solve biocompatibility issues and excessive
accumulation in tissues.

Usually biomedical applications require long-circulating delivery vectors
with constant release for days\cite{Murray}. This implies that polymer
degradation in such systems is much slower than the time required to reach
the thermodynamic equilibrium. Kinetics of micelle formation from monomers
can be rather fast. Characteristic times of exchange of oligomers between
micelles and the bulk and the relaxation do not exceed milliseconds \cite%
{Kaatze}, while for longer block copolymers, the characteristic time is of
order of minutes\cite{Lund,Lund1}. Thus, in such systems the self-assembly
in micelles is a quasi-equilibrium process with a fixed length distribution
of the blocks, that changes with time. Micelles of degradable block
copolymers can assemble and reassemble changing their internal structure and
shape\cite{Discher,Hu} thus changing the loading capacity of the cores. The
balance between steric repulsion of hydrophilic blocks in the coronas of
micelles and interaction of hydrophobic blocks forming part of the cores
defines the finite size of micelles. Degradation of hydrophilic blocks\cite%
{Frazza,Torchilin1} would then lead to destabilization of micelles leading
to fusion and formation of larger cores. In turn, degradation of hydrophobic
blocks in the cores\cite{Lavasanifar,Avgoustakis,Sun,Tew,Wang,Cho,Li} may
induce splitting the cores of micelles.

\begin{figure*}
\begin{center}
\includegraphics[width=12cm]{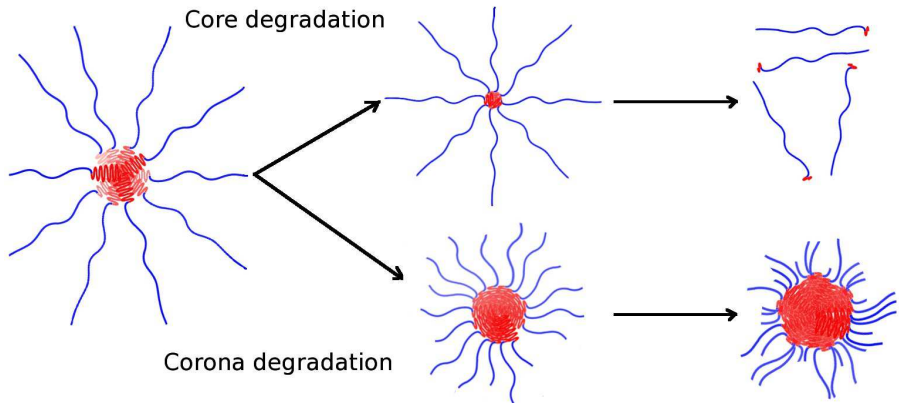}
\end{center}
\caption{Quasi-equilibrium degradation of micelles. Core degradation induces
the disassembly of micelles while corona degradation induces growth of
micelles.}
\label{Fig:pic}
\end{figure*}

Using a scaling theory we study the interplay between the degradation of one
of the blocks of block copolymers and equilibrium self-assembly and
re-assembly of block copolymers into micelles in case of both corona and
core degradation. We present a scaling theory of block-copolymer
self-assembly into micelles\cite{Sens,Baulin}, and discuss the degradation
kinetics of the polymer blocks. However, we note that scaling theories have
limitations, (i) Scaling theories in principle only apply to (infinitely)
long chains. (ii) the theory does not include kinetics; (iii) it does not
include morphological changes. We assume two types of degradation
mechanisms, (i) random chain scission mechanism\cite{Krapivsky}, which
implies stochastic process of a polymer chain division at random position.
This mechanism causes a gradual decrease in the number average molecular
weight. Such random division of a chain is usually observed in hydrolytic
degradation of polyesters\cite{Li1,Smith} or certain polyamides\cite%
{Galbis,Galan}. (ii) End-evaporation mechanism\cite{Marques,Schoot}, when
the monomers are gradually detached from the ends of the polymer chain. This
mechanism is typical for enzymatic degradation. The degradation of the
blocks shifts the critical micellar concentration (CMC) and changes the size
distributions and the average sizes of the micelles (Figure %
\ref{Fig:pic}). The degradation of the chains can induce or suppress
the self-assembly process.

\section{Self-assembly}

Diblock copolymers composed of soluble and insoluble blocks can
spontaneously self-assemble into micelles\cite{Hamley}. The micellization is
the entropy driven process, the entropy of mixture of individual
chains in the solution is balanced by the tendency of insoluble blocks to reduce contacts with the
solvent in the cores of micelles. Thus, the stability of the micelles is
defined by the energy of insoluble blocks forming
compact cores and the steric repulsion of soluble blocks in the coronas of
micelles\cite{Sens,Baulin}. Characteristic equilibration time of block
copolymers depends exponentially on the length of the insoluble block\cite%
{Lund,Lund1}. In the cases of long insoluble blocks and high interfacial
tensions\cite{Richter} the relaxation time may be hours, and thus
non-equilibrium kinetics should be considered. However, we focus on
situations when the hydrophobic block is relatively short or the interfacial
tension is low. In this case the relaxation time is small (milliseconds)\cite%
{Kaatze} and thus the kinetics of self-assembly is much faster than the
degradation. If the degradation process is slow enough to allow for the
equilibrium assembly of block copolymers into micelles, the scaling model of
equilibrium self-assembly can be applied. We assume that only one block,
either insoluble or soluble, is degradable, so that the total number of
block copolymers does not change with time. In addition, we consider
spherical micelles, while morphological transitions are not considered.
We denote $c_{p}$ the number density of aggregates comprising of $p$ copolymers, where $p$ is the aggregation number.
The total free energy per unit volume of the solution of copolymers at a given
time is

\begin{equation}
\frac{F}{kT}=\sum_{p=1}^{\infty }c_{p}[\ln (\frac{c_{p}}{e}\Lambda ^{3})+%
\frac{F_{p}}{kT}]+\int_{0}^{\infty }c(n)\ln (\frac{c(n)}{e}\Lambda ^{3})dn
\label{F}
\end{equation}%
where the first term is the entropy and $F_{p}$ is the free energy of a
micelle comprising $p$ copolymers, $\Lambda $ is the de Broglie wavelength. $%
p=1$ corresponds to individual block copolymers contributions, while the
entropy of free fragments is taken into account in the last term.
Degradation of blocks provokes the detachment of fragments of different
lengths floating in the solution and the last term takes into account the
entropy of fragments where $c(n)$ is their length distribution function. The
total number of monomers in the self-assembly and the degradation process is
conserved. This is reflected in the conservation of mass condition,

\begin{equation}
\sum_{p=1}^{\infty }pc_{p}=\varphi  \label{phi}
\end{equation}%
where $\varphi $ is the total copolymer concentration. Since we consider the
degradation of one block, the total number of copolymers in the solution is
not changed with time and the degradation does not affect this condition.
Minimization of the free energy (\ref{F}) subject to the constraint (\ref%
{phi}) gives the quasi-equilibrium distribution of the copolymers in the
micelles \cite{Baulin2},

\begin{equation}
c_{p}=c_{1}^{p}\Lambda ^{3(p-1)}\exp (-\frac{F_{p}-pF_{1}}{kT})  \label{cp}
\end{equation}%
This expression describes the distribution of micelles of degradable
copolymers at each time.

Explicit form of the free energy $F_{p}$ is defined by the molecular
structure of copolymers forming a micelle and is the sum of the corona and
the core contributions,

\begin{equation}
F_{p}=F_{p}^{corona}+F_{p}^{core}  \label{Fp}
\end{equation}

The free energy of individual chains, $p=1$, is also described by this
expression, where the corona term transforms in the entropy contribution of
a linear chain and the insoluble block gives the corresponding core
contribution. The exact expressions of $F_{p}^{corona}$ and $F_{p}^{core}$
depend on degradation mechanism and are functions of time.

In the following, we denote the chain length of the soluble block as $N$ and
the chain length of the insoluble block as $N_{c}$. We assume sufficiently
long soluble chains, $N\gg 1$, and the monomers of both blocks being of the
same size $a$.

\section{Degradation kinetics}

We consider random chain scission and end evaporation mechanisms and
degradation of soluble and insoluble blocks.

\begin{figure*}[th]
\begin{center}
\includegraphics[width=12cm]{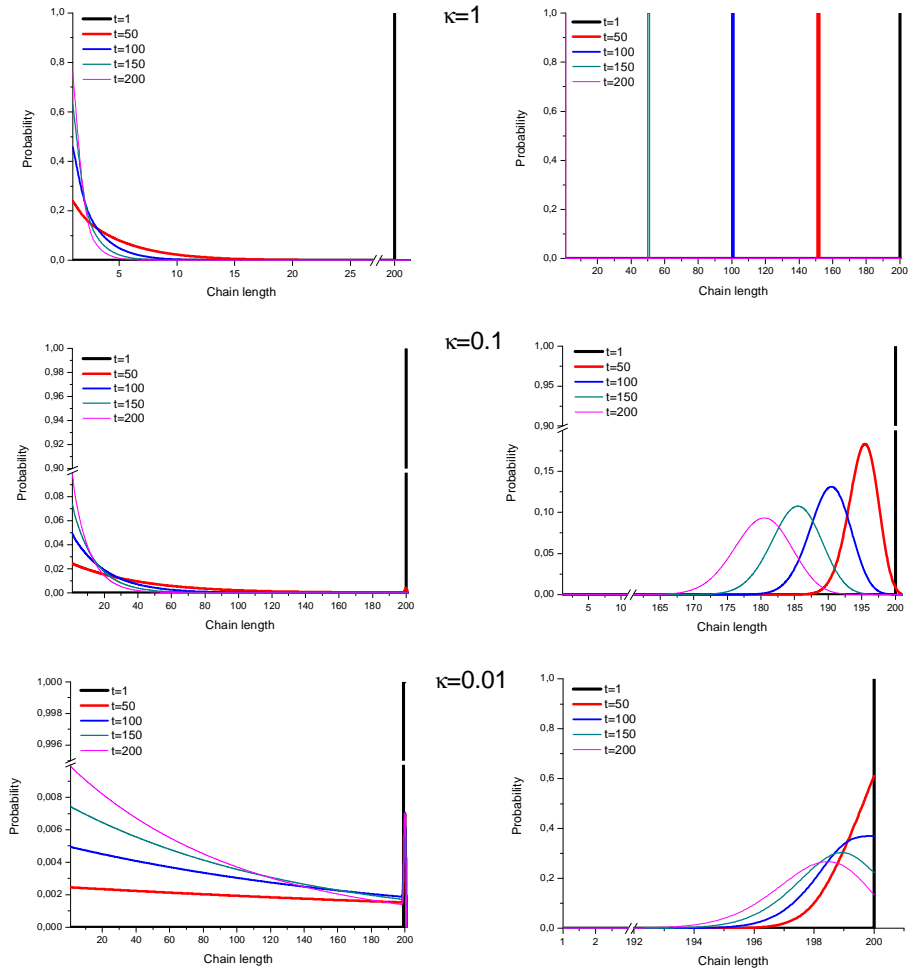}
\end{center}
\caption{Length distributions in the random scission (left column) and the
end evaporation (right column) mechanisms for different degradation rates $%
\protect\kappa $ and initial length $200$.}
\label{Fig:kin}
\end{figure*}

\subsection{1) Random chain scission}

Random scission mechanism implies homogeneous distribution of splitting
points along the chain. At a given time a chain is divided into random
parts. Thus, the distribution of fragments of different lengths $P(n,t)$ is
given by\cite{Krapivsky}
\begin{equation}
\frac{1}{\kappa }\frac{\partial P(n,t)}{\partial t}=-nP(n,t)+\int_{n}^{%
\infty }dyP(y,t)  \label{Degr}
\end{equation}%
The negative term refers to the loss of chains of length $n$, and the
positive term describes the gain of chains of length $n$ due to the
degradation of longer chains (with lengths more than $n$) at a given time $t$%
. The coefficient $\kappa $ takes into account the fraction of chains that
degrade simultaneously at each time, practically defining the timescale.
Initial distribution of chains at $t=0$ is assumed to be monodisperse. This
equation is solved numerically by considering the integral in the right hand
side $Q(n,t)=\int_{n}^{\infty }dyP(y,t)$ a numerical function. This function
is calculated for a given time as a function of $n$ and then the equation (%
\ref{Degr}) is solved numerically. Typical length distribution functions are
shown in Figure \ref{Fig:kin}, left column. Initial homogeneous
distribution gradually disperses and shifts to small $n$.

\subsubsection{a) Core degradation}

Insoluble blocks tend to avoid contacts with water and favor assembly into
the core of the micelle. The degradation of insoluble blocks makes block
copolymers more soluble and this would shift the equilibrium towards smaller
micelles. Assuming that the core of a micelle is a dense and homogeneous
sphere formed by $p$ copolymers, the free energy of the core is given by\cite%
{Baulin2}

\begin{figure}[th]
\begin{center}
\includegraphics[width=7cm]{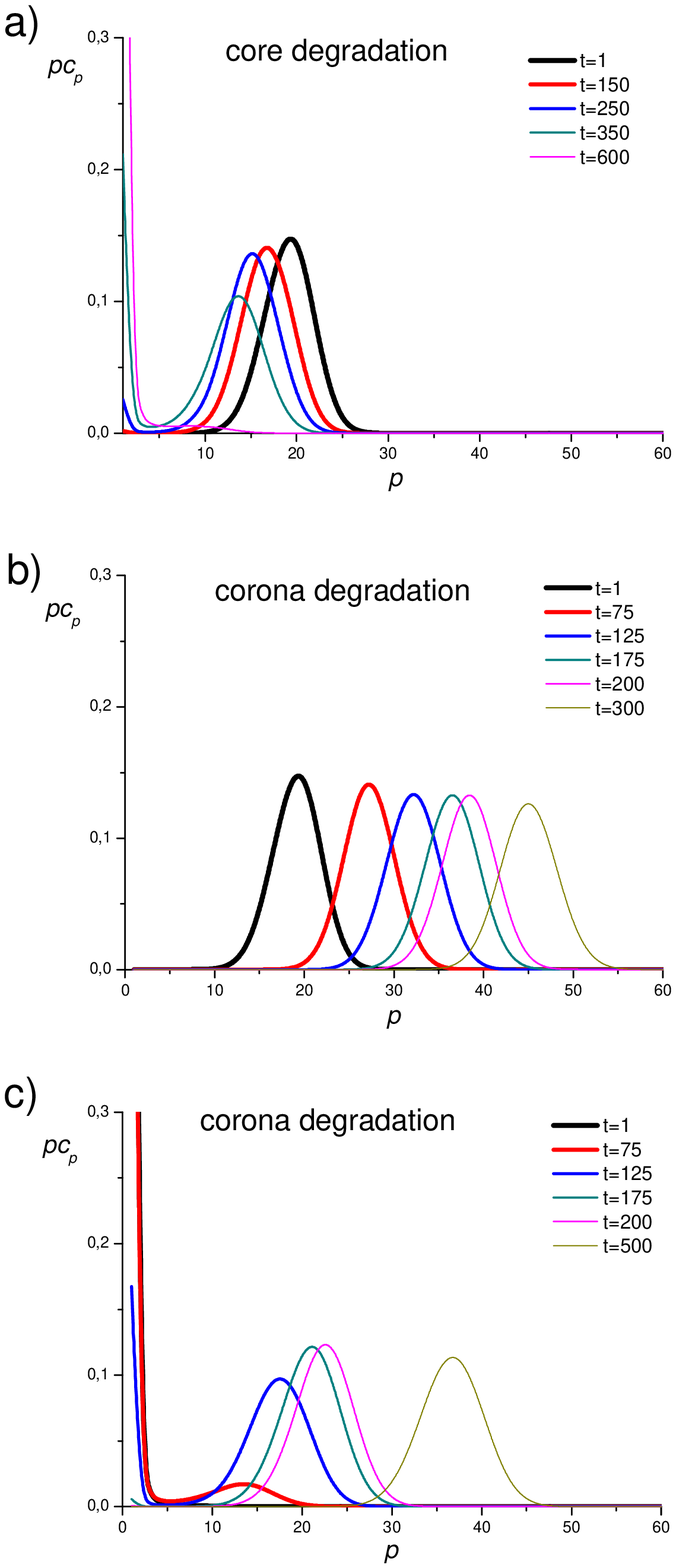}
\end{center}
\caption{ Time evolution of the normalized size distribution $pc_{p}$. a)
core degradation above CMC, $\protect\sigma =0.81$, $\protect\kappa =0.0003$
b) corona degradation above CMC, $\protect\sigma =0.81$, $\protect\kappa %
=0.01$ c) corona degradation below CMC, $\protect\sigma =0.75$, $\protect%
\kappa =0.01$. The initial lengths of the blocks are: $N=200$, $N_{c}=20$,
copolymer concentration $c_{1}=10^{-6}$.}
\label{Fig:cp}
\end{figure}

\begin{equation}
F_{p}^{core}=kT(36\pi )^{\frac{1}{3}}\sigma N_{c}^{\frac{2}{3}}p^{\frac{2}{3}%
}+\frac{3\pi ^{2}p^{\frac{5}{3}}}{80N_{c}^{\frac{1}{3}}}  \label{Fcore}
\end{equation}%
where $\sigma $ is the surface tension of a sphere of radius $R_{c}=(3/(4\pi
)pN_{c})^{1/3}a$ and the second term describes the elastic contributions
arising from the stretching of insoluble blocks in the core\cite{Zhulina}.
The "effective" surface tension of the core $\sigma $ describes implicitly
the fact that the insoluble chains tend to avoid contact with the solvent by
forming a dense core. The lengths of the insoluble blocks $N_{c}$ depend on
time and for each time $t$ the length distribution is given by equation of
random scission degradation (\ref{Degr}).

Steric repulsion in the corona of the micelles, formed by soluble blocks,
penalizes the formation of large micelles. Since the soluble block does not
degrade, the corona of micelles is composed of chains of equal length $N$.
The partition function of a monodisperse star $Z_{p}$ yields in the form\cite%
{Johner,Baulin1}

\begin{equation}
Z_{p}\propto N^{\gamma _{p}-1}
\end{equation}%
where $N$ is the length of the arm\cite{Baulin1}. $\gamma _{p}$ are the
universal critical exponents of the star polymers\cite{Duplantier}. The
numerical values of $\gamma _{p}$ are known exactly for a wide range of $p$
\cite{Grassberger} and in the range $0<p<200$ can be interpolated by the
power law expression $\gamma _{p}=1-0.0893(p-1.5)^{1.68}$. With this, the
free energy of polydisperse corona is given by

\begin{equation}
F_{p}^{corona}=-kT\ln Z_{p}=-kT(\gamma _{p}-1)\ln N  \label{Fcorona}
\end{equation}%
Eqs. (\ref{Fcore}) and (\ref{Fcorona}) define the free energy of micelles of
$p$ copolymers and the free energy of individual chains for $p=1$. It allows
to calculate the size distribution of micelles as a function of time (\ref%
{cp}). The results are present in Figure \ref{Fig:cp}a) The degradation
of the core starts when the micelles are formed (concentration above CMC).
The micelles gradually decrease in size and disassemble. The degradation
rate $\kappa $ is related to the time step in the degradation equation (\ref%
{Degr}). The chosen value $\kappa =0.0003$ is low enough to insure gradual
changes in the size of the micelles. if the rate is higher, the micelles
would disappear faster. In a real experimental situation this parameter
connects the time step with real time, e.g. $\kappa =0.0003$ signify that
only three of ten thousand chains disassociate at one time step. In
experiments this parameter may vary in a wide range.

\subsubsection{b) Corona degradation}

Situation is different when the soluble blocks can degrade while the
insoluble blocks are stable. The cores of the micelles are formed by
insoluble blocks and the core contribution has the same form as in previous
case, eq. (\ref{Fcore}), but $N_{c}$ remains constant. In turn, soluble blocks
forming corona now can degrade with time and the corona contribution
changes. Coronas of micelles are formed by polydisperse arms with the length
distribution $P(n,t)$ given by (\ref{Degr}). The partition function of a
polydisperse star $Z_{p}$ yields in the form\cite{Johner,Baulin1}

\begin{figure}[th]
\begin{center}
\includegraphics[width=8.25cm]{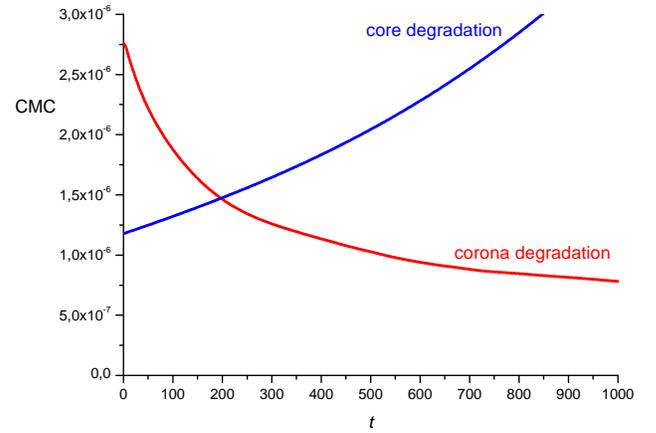}
\end{center}
\caption{Time dependence of the critical micelle concentration (CMC).
Parameters are the same as in Figure \ref{Fig:cp}a) and b).}
\label{Fig:CMC}
\end{figure}

\begin{equation}
Z_{p}\propto n_{1}^{\gamma _{p}-\gamma _{p-1}}n_{2}^{\gamma _{p-1}-\gamma
_{p-2}}\ldots n_{p-1}^{\gamma _{2}-\gamma _{1}}n_{p}^{\gamma _{1}-1}
\end{equation}%
where $n_{1}<n_{2}<\ldots <n_{p-1}<n_{p}$ are the lengths of the
corresponding arms, sorted in ascending order\cite{Baulin1}. This partition
function leads to the corresponding expression of the free energy of the
corona,

\begin{eqnarray}
F_{p}^{corona}&=&-kT\left[ (\gamma _{p}-\gamma _{p-1})\ln n_{1}+(\gamma
_{p-1}-\right.\\ \nonumber
&& \left. \gamma _{p-2})\ln n_{2}+\ldots+(\gamma _{1}-1)\ln n_{p}\right]
\end{eqnarray}

This expression defines, together with eq. (\ref{Fcore}), the free energy of the
micelles and individual chains in the solution (\ref{Fp}). In fact, the scaling
expression of the corona contribution of a micelle with one arm,
$F_{p=1}^{corona}=-kT(\gamma _{1}-1)\ln N$ corresponds
exactly to the scaling expression of a soluble chain in a solution\cite{Duplantier}.

The resulting
size distribution of micelles (\ref{cp}) is shown in Figure \ref{Fig:cp}%
b) and c) for different times. The effect of corona degradation is opposite
to the case of core degradation. Figure \ref{Fig:cp}b) shows the corona
degradation kinetics for the same initial conditions as in Figure %
\ref{Fig:cp}a) when the micelles are formed for conditions above CMC.
Degradation of corona induces self-assembly of micelles from initially
homogeneous solution of individual chains. Individual chains associate into
micelles and the distribution of micelles grows. Consequent degradation
leads to the shrinkage of the corona, thus larger micelles become more
stable and small micelles unstable. The size distribution of the micelles
becomes broader with time due to increased polydispersity. Further
degradation of hydrophilic blocks would lead to the growth of micelles,
morphological changes\cite{Discher} of the micelle shape and consequent bulk
phase separation.

Another important effect of corona degradation is the possibility to induce
self-assembly of micelles from the solution below CMC (Figure %
\ref{Fig:cp}c)) The degradation induces formation of micelles from
initially homogeneous solution first for small aggregation numbers.
Consequent degradation increases the number of micelles, the polydispersity
and the average size. Thus, the degradation influence the CMC, which is time
dependent, Figure \ref{Fig:CMC}. It decreases with time in case of
corona degradation (red curve) and increases with time in case of core
degradation (blue curve).

\begin{figure}[th]
\begin{center}
\includegraphics[width=8.25cm]{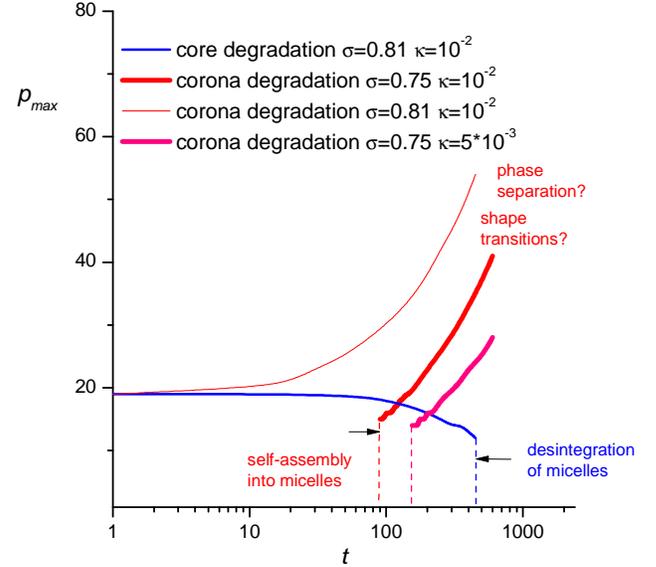}
\end{center}
\caption{Time evolution of the aggregation number of the maximum of the size
distribution, $p_{m}ax$. Parameters are the same as in \protect\ref{Fig:cp}.}
\label{Fig:pmax}
\end{figure}

The effect of degradation on micellization process can be summarized in the
plot showing the position of the maximum $p_{max }$ of the size distribution
$c_{p}$, Figure \ref{Fig:pmax}. In case of core degradation (blue line)
the maximum of the distribution moves to small numbers (see Figure %
\ref{Fig:cp}a)) until the micelles disappear completely (dashed line),
while in case of corona degradation (red curve) the size of the micelles
increases (see Figure \ref{Fig:cp}b)) until the morphology changes or
phase separation occur. If the degradation starts below CMC, the
self-assembly into micelles starts at a given time which depends on the rate
of degradation $\kappa $.

\subsection{2) End evaporation}

Random division of a chain is not the only degradation mechanism of polymer
chains. Degradation in certain chemical reactions and enzymatic degradation
may lead to gradual decrease of the chains from the ends
(chain-end-activated degradation). The process of loosing the monomers from
the end is described by the following process, $P(n,t+1)-P(n,t)=\kappa
(P(n+1,t)-P(n,t))$ which can be written in the integral form similar to (\ref%
{Degr}) as

\begin{figure}[th]
\begin{center}
\includegraphics[width=8.25cm]{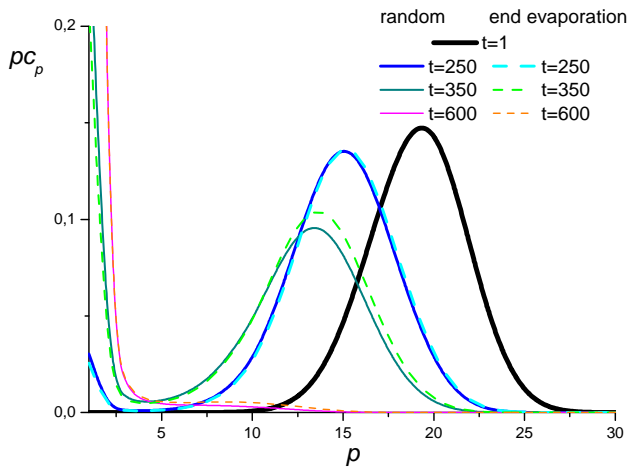}
\end{center}
\caption{End evaporation kinetics of core degradation in comparison with
random scission. Parameters are the same as in Figure \ref{Fig:cp}a)
except the degradation rate, $\protect\kappa =0.03$.}
\label{Fig:cp1}
\end{figure}

\begin{equation}
\frac{1}{\kappa }\frac{\partial P(n,t)}{\partial t}=-P(n,t)+\int_{n}^{\infty
}\delta (y-n-1)dyP(y,t)
\end{equation}%
where $\delta $ is the Dirac function. Using this equation instead of (\ref%
{Degr}), one can obtain the time dependence of the free energy of micelles
for core and corona degradation. Figure \ref{Fig:kin} right column
presents the kinetics end evaporation degradation. The starting and the
final chain length distributions are very close for both types of
degradation. That is why the micelle distributions shown for core
degradation in Figure \ref{Fig:cp1} also coincide. The difference is
only seen for intermediate times. However, end evaporation degradation is
much slower than random scission, thus the rates of degradation differ in
this example 100 times.

In conclusion, scaling theory of quasi-equilibrium micellization coupled
with the degradation of the blocks demonstrates that the degradation of
hydrophilic blocks can induce self-assembly of copolymers into micelles and
increase the size of the micelles, while the degradation of the hydrophobic
blocks destabilize the micelles, reduce the equilibrium size of the micelles
and can lead to complete disassociation of the micelles. It is valid for
random scission mechanism assumed for degradation mechanics as well as for
enzymatic (chain-end) scission mechanism. These findings may suggest the
ways of controlled self-assembly and destabilization of micelles by
degradation of the blocks. Our model do not account for morphological
transitions and we plan to study them in future with a more detailed
microscopic theory\cite{Pogodin}.

\acknowledgments{The authors acknowledge financial help from Spanish Ministry of
education MICINN via project CTQ2008-06469/PPQ.}


\providecommand*\mcitethebibliography{\thebibliography}
\csname @ifundefined\endcsname{endmcitethebibliography}
  {\let\endmcitethebibliography\endthebibliography}{}

\end{document}